\documentstyle[aps,prb,psfig,epsfig]{revtex}

\newcommand \be{\begin{eqnarray}}
\newcommand \ee{\end{eqnarray}}
                                               
\begin{document}
\draft
\twocolumn[\hsize\textwidth\columnwidth\hsize 
           \csname @twocolumnfalse\endcsname

\title{Dynamical local field, compressibility and frequency sum rules
  for quasiparticles}
\author{Klaus Morawetz}
\address{Max-Planck-Institute for the Physics of Complex Systems, 
Noethnitzer Str. 38, 01187 Dresden, Germany}
\maketitle
\date{\today}
\maketitle
\begin{abstract}
The finite temperature 
dynamical response function including the dynamical local field is
derived within a quasiparticle picture for interacting one-, two- and
three 
dimensional Fermi systems.
The correlations are assumed to be given by a density dependent effective mass,
quasiparticle energy shift and relaxation time. The latter one describes
disorder or collisional effects. This parameterization of correlations
includes local density 
functionals as a special case and is therefore applicable for density functional theories. 
With a single static local field, the third order
frequency sum rule can be fulfilled simultaneously with the
compressibility sum rule by relating the effective mass and
quasiparticle energy
shift to
the structure function or pair correlation function. Consequently, solely local
density functionals without taking into account effective masses
cannot fulfill both
sum rules simultaneously with a static local field. The comparison to
the Monte-Carlo data seems to support such quasiparticle picture.
\end{abstract}
\vskip2pc]

\section{Introduction}

The response of an interacting Fermi system with the
potential $V_q$ to an external
perturbation is the basic source of our knowledge about the interaction
and dynamical as well as statical properties of the
system. This response function has been therefore a central issue of
many body theories. 

The density response function gives the variation of the density in
terms of the external potential
\be
\delta n(q,\omega)=\chi(q,\omega) V^{\rm ext}(q,\omega).
\label{dela}
\ee 
The
polarization is defined as the density variation in terms of the
induced potential
\be
\delta n(q,\omega)=\Pi(q,\omega) \delta V^{\rm ind}(q,\omega)
\ee 
where we suppress the notation of obvious $q$-dependence in the following.
The induced potential itself is
the sum of
the external potential and the effective interaction
potential $(V_q+f_q(\omega)) \delta n$ 
\be
\delta V^{\rm ind}(\omega)=(V_q+f_q(\omega)) \delta n(\omega)+V^{\rm
  ext}(\omega).
\label{dela1}
\ee 
Therefore, from (\ref{dela})-(\ref{dela1}) we have
the relation between response and polarization
\be
\chi(\omega)={\Pi(\omega)\over 1-(V_q+f_q(\omega)) \Pi(\omega)}.
\label{chim}
\ee
The local field $f_q(\omega)$ describes the modification in the restoring force
brought about by particle correlations. This field prevents the
particles from sampling the full effect of interaction at short
distances \cite{IKP84}.

The dielectric function relates now the induced densities to the
external potential via
\be
{1 \over \epsilon(\omega)}=1+{V_q \delta n(\omega) \over V^{\rm ext}(\omega)}\equiv 1+ V_q
\chi(\omega)
\ee
such that the dielectric function reads
\be
\epsilon(\omega)=1-{V_q \Pi(\omega) \over 1-f_q(\omega) \Pi(\omega)}.
\label{deele}
\ee

The theoretical effort consists in determining the local field $f_q(\omega)$ which
represents
the local correlation and which depleted the induced potential by 
$f_q(\omega)=-G(\omega) V_q$.
As long as this local field is a dynamical one this is an exact
relation. The different theoretical treatments differ in 
this local field corrections, for an overview see
\cite{KKM89}. Mostly static approximations,
$f_q\equiv f_q(0)=-V_q G$,
have been proposed in the past. It has started with the pioneering
work of Hubbard
\cite{H57} who first introduced the notation of local field and 
took into account the exchange--hole correction resulting in
\be
G_H&=&\frac 1 2 {q^2 \over q^2 +k_f^2}=\left \{\matrix{
\frac 1 2 {q^2 \over k_f^2} +o(q^3)\cr
\frac 1 2 +o(1/q^2) }\right ..
\ee
While this expression has established a remarkable improvement of the
dielectric function in
random phase approximation (RPA), 
it has been soon recognized insufficient due to
the lack of self-consistency which leads the
pair correlation function still to unphysical negative
values. This has been repaired by Singwi et. al. \cite{STLS68} by
using exchange-correlations
\be
G_{STLS}&=&-\frac 1 n \int {d k \over (2 \pi)^3} {(k\cdot q)\over k^2}
  (S_{k-q}-1)
\nonumber\\
&=&
\left \{\matrix{\gamma {q^2 \over k_f^2} +o(q^3)\cr
1-g_0 +o(1/q^2) }\right .
\label{STLS}
\ee
with 
\be
\gamma=-{1\over s k_f} \int\limits_0^\infty d q (S_q-1)
\label{gamma}
\ee 
where $s$ is the spin degeneracy and where the static structure factor
\be
S_q&=&-\int {d \omega \over n \pi V_q} {{\rm Im}\epsilon^{-1}(\omega) \over 1-{\rm
      e}^{-\beta \omega}}
\nonumber\\
&=&\int {d \omega \over n \pi} {1 \over 1-{\rm
      e}^{-\beta \omega}}{\rm Im}{\Pi(\omega)\over 1-(f_q(\omega)+V_q) \Pi(\omega)}
\label{S}
\ee
with inverse temperature $\beta=1/T$ is linked to the pair correlation
function via
\be
g_r-1=\frac 1 n \int {dq \over (2 \pi)^3} {\rm e}^{i q r} (S_q-1).
\label{gs}
\ee
This provides a self-consistent problem in solving dielectric function,
structure function and static local field simultaneously. The advantages of
this result compared to  the Hubbard result with respect to the
pair correlation function and large wavevector limit has been discussed by
\cite{S70}. Recent comparisons with molecular dynamics simulations for a
hard sphere gas is presented in \cite{SSF96} where a good agreement is found for
thermodynamical properties. 

The expression (\ref{STLS}) has been improved further by Pathak and Vashishta
\cite{PV73}
demanding that the
response function should fulfill the third order frequency sum rule
\cite{P65} which resulted into
\be
G_{PV}&=&-\frac 1 n \int {d k \over (2 \pi)^3} {(k\cdot q)^2\over q^4}{V_k
    \over V_q}
  (S_{k-q}-S_k)
\nonumber\\
&=&
\left \{\matrix{\frac 2 5 \gamma {q^2 \over k_f^2} +o(q^3)\cr
\frac 2 3 (1-g_0) +o(1/q^2) }\right .
\ee
leading to the improved small distance limit discussed in
\cite{N74}. The difference at short distance to $G_{STLS}$
comes from the motion of particles inside the correlation hole which is
condensed in the dynamical behavior \cite{K73}.

If one takes into account the difference between uncorrelated and
correlated kinetic energy \cite{VG73,IKP84} one obtains an additional
$-\Delta E=-{2\over n^2 V_q} (E_{int}-E)$ term to $G_{PV}$. This 
comes from the difference in correlated and uncorrelated 
occupation numbers which can be expressed by a coupling constant
integration and can be linked via the virial theorem to density
derivatives of the pair correlation function \cite{DA86,RA94}. 

Parallel to the above discussions there has been different
improvements to derive local fields from the virial formula 
\cite{N74,VG73,VS72} which have resulted into expressions known from density variations 
\be
G_{VS}=(1+a n {\partial \over \partial n} ) G_{STLS}
\ee
with various $1/2 \le a \le 1$, see \cite{VS72}. This procedure satisfy
the compressibility sum rule almost exactly.

In \cite{VG73} it was shown for Coulomb systems that
one cannot construct a static local field factor
which fulfills both limits, the compressibility 
and
the third order sum rule (\ref{req}) since it would violate the
theorem of Ferrell, $d^2 E_0/d(e^2)^2\le0$.
The same conclusion are obtained in \cite{I84} using the
virial theorem. 

Therefore
the concentration is now mostly focussed on the construction of
dynamical local field corrections \cite{RA94,SchB93,RRWR99,RRRW00}. 
The quantum versions of the Singwi-Tosi-Land-Sj{\"o}lander, $G_{STLS}$,
and Vahishta-Singwi, $G_{VS}$, theories have been discussed in 
\cite{SchB93,HR87,DT99}. These lead to positive values for the pair 
distribution function at short distances valid for rather low densities
\cite{SchB93}.
 There it has been focussed on dynamical properties of the dynamic
local field. While the high frequency limit is monotonic and similar
to STLS and VS, the static limit can even exhibit peaked structures
which can give rise to charge density waves 
underlying the nontrivial character of dynamical
behavior. Unfortunately even the dynamic quantum
version of the
Singwi-Tosi-Land-Sj{\"o}lander local field cannot fulfill the
compressibility sum rule \cite{TB99}. We will show here that
from a
dynamical local field one can derive a static local field fulfilling
both sum rules simultaneously if one takes into account the effective mass. 
This will resolve the puzzle of sum rules. 

Recent improvements of the response function are basically due to
numerical studies of Monte Carlo \cite{BSA94,OB94,MCS95,G97} or 
molecular dynamical simulations 
\cite{PRBB85,SSF96}. An interesting first principle numerical scheme
is to solve the time dependent Kadanoff and Baym equations including
an external field \cite{KB00}.
 Due to the variation of internal lines, already a Born
diagram leads to a linear response which includes high order 
vortex corrections fulfilling sum rules consistently. 
The third order frequency sum rule regains importance
for reduced dimensional layered structures \cite{G97,GL92,CG97}.
All results in this paper here can be straight forwardly generalized also
to reduced dimensions as given in 
appendix~\ref{a} for one, two and three dimensions. This could have an
impact on recent
discussions of two-layered electron gasses \cite{CD97,CD91,TD96}.

Here we want to return to the analytical investigations and will
show that there exist a possibility to
fulfill with one static local field correction both
requirements, the third order sum rule and the compressibility sum
rule. This is performed by working within a quasiparticle
picture determining the effective mass appropriately. Within the
frame of the quasiparticle picture we will derive an explicit expression
for the dynamical local field factor
which leads to the desired static limits. We obtain the identity
\be
G&=&G_{PV}+{2 \over n^2 V_q} (E-E_{int})\nonumber\\
&\equiv&-{1\over V_q}\left ({\partial \Delta \over \partial n}-{1\over 2 n
    m} {\partial \ln m \over \partial \ln n} \left ({\Pi_2(0)\over
      \Pi_0(0)}-{q^2 \over 4}\right )\right )
\ee
with moments of the polarization given in appendix~\ref{a} and the effective mass
$m(n)$ and selfenergy $\Delta(n)$.
It will provide a recipe how
to construct a quasiparticle picture by the knowledge of the structure
factor at small distances from experiments or simulations \cite{PRBB85,BSA94,SSF96}. 
This in turn leads to an easy microscopic parameterization in terms of
the effective
mass and quasiparticle energy shift which could be compared directly
to microscopic theories. 

The underlying principle is analogous to the one found in literature 
\cite{NI89a,NI89b} where a response function was parameterized 
explicitly fulfilling sum rules and different constraints. Other
parameterizations can be found in \cite{RA94} from variational
approaches which are exact in the high density limit. A different line of
constructing the response function uses the frequency moments  
resulting in recurrence relations \cite{H82,HL93}.

Here in this paper we will give an alternative approach which uses
general parameterizations of the selfenergy in terms of a functional
which might depend on the density, energy and current. We restrict here
to a one component system though the generalization to multicomponent
systems is straight forward \cite{MWF97,MFW99} and considered in
different approaches \cite{IMTY85,KG90,Mor00}.

In the next chapter we review shortly the compressibility sum rule and
the third order frequency sum rule. In chapter III we give the
dynamical response for quasiparticles which is a special case of the
general structure derived earlier \cite{MF99}. We show that the
correct compressibility appears and the third order sum rule can be
satisfied if the effective mass and quasiparticle energy is chosen
appropriately. Chapter IV will present some numerical results on the
unpolarized electron gas at zero temperature and the comparison with
Monte Carlo simulations are discussed. Chapter V summarizes the results and in appendix~\ref{a} we
give frequently occurring correlation functions and practical forms for
calculation in one-, two- and three dimensions. While all
formulas in the main text are written as three dimensional, they hold
for one- and two dimensions as well. Only the explicit correlation functions
in appendix~\ref{a} have to be used as outlined
there. Also the often required long wavelength expansion of these
correlation functions are given in
appendix~\ref{expans}. Appendix~\ref{pert} finally 
is devoted to
the short sketch of perturbation
theory and the derivation of the used sum rules for one, two and three
dimensions. 

\section{Determination of static local field factor}

Let us discuss two different boundaries for the static limit of
the local field. This will be the compressibility and the third order
frequency sum rule. 

\subsection{Compressibility}

First we have to know how the
compressibility should look like. This is particularly simple in the
quasiparticle picture which we will use. In the quasiparticle picture 
the one--particle distribution function is a Fermi distribution
\be
F(p)=({\rm e}^{\beta ({p^2 \over 2 m}+\Delta-\mu)}+1)^{-1}
\label{F}
\ee
where the density dependent effective mass $m$ and the selfenergy
shift $\Delta$ are obtained either 
from microscopic calculations or, as proposed here, from the sum
rules. 
Thorough the paper we will understand now the masses as
effective masses.

From microscopic approaches the effective selfenergy shift and the
effective mass are coming from the knowledge of the selfenergy
$\sigma(p,\omega)$ which determines the quasiparticle energy
$\epsilon$  via
\be
\epsilon={p^2\over 2 m_0}+\sigma(p,\epsilon).
\ee
The velocity of the quasiparticles is given by $\partial
\epsilon/\partial p$ which leads to the definition of the effective mass
\be
{1\over m}=\left .{{1\over m_0}+{1\over p}\partial_p\sigma \over 1-\partial_\omega
  \sigma}\right |_{\omega=\epsilon_f,p=p_f}
\label{mass1}
\ee
where the momenta and energy are set to the Fermi momenta and energy
after derivatives. Consequently the quasiparticle energy can be
approximated by
$\epsilon\approx{p^2\over 2 m}+\Delta$ with the
effective mass (\ref{mass1}) and the energy shift
$\Delta=\sigma(p_f,\epsilon_f)$. This approximation has to be replaced
by thermal averaging when finite temperature systems are considered.
A useful method would also be to read off the quasiparticle
parameterization from current parameterizations of the momentum
distribution \cite{Z02,GZ02}.

Using the definition of the compressibility one obtains directly from
(\ref{F}) 
\be
{\cal K}&=&{1\over n^2} {\partial n\over \partial \mu}=-{\beta \over n^2} \int
{d p \over (2 \pi)^3} F(p)(1-F(p))
\nonumber\\&&\qquad\qquad \times
 \left ([\Delta'(n) -{p^2 \over 2 m} (\ln
m)'] {\partial n \over \partial \mu}-1\right )
\nonumber\\
&&=
{{\cal K}_0\over 1+n^2  {\delta \over \delta n} \Delta {\cal K}_0-\frac 3 2
  {\partial \ln m \over \partial \ln n}}
\label{direct}
\ee
where the free compressibility is ${\cal K}_0={\beta \over n^2} \int
{d p \over (2 \pi)^3} F(p)(1-F(p))$. Alternatively, the energy shift
$\Delta$ and the effective mass can be expressed by Landau parameter.

In the next chapter we will present a consistent dynamical response
function for the quasiparticle picture such that the correlations are
parameterized by quasiparticles with an effective mass, 
an energy shift and a relaxation time. From this we will obtain the correct
compressibility (\ref{direct}) from the dynamical response via the
static limit obeying the frequency sum rules. 
Actually, a static local field can be constructed provided we choose the
effective mass appropriately. This will lead to a recipe how the
effective mass
can be determined from the structure factor which is well known from
Monte-Carlo simulations or experiments.

The conventional compressibility sum rule \cite{PNO68} reads with
$\lim\limits_{q\to 0}\Pi=-n^2 {\cal K}_0$ and (\ref{deele})
\be
-\lim\limits_{q\to 0}{2\over \pi V_q} \int\limits_0^\infty {d\omega' \over \omega'}{\rm Im}
\,\epsilon(\omega') &=&\lim\limits_{q\to 0}{\rm Re}
  \, {\Pi(0,q)\over 1-f_q(0) \Pi(0,q)}\nonumber\\
&=&-\lim\limits_{q\to 0}{n^2 {\cal K}_0\over 1+f_q(0) n^2 {\cal K}_0}\nonumber\\
&\equiv& -n^2 {\cal K}
\label{compress}
\ee
such that we can expect from the correct result (\ref{direct})
that the static local field has the form
\be
\lim\limits_{q\to 0}f_q(0)={\partial \Delta\over \partial n} -{3
  \over 2 n^2 {\cal K}_0} {\partial \ln m \over \partial \ln n}.
\label{fexp}
\ee
We will present a dynamical local field which leads in the static
limit exactly to this desired result (\ref{fexp}).

\subsection{Frequency sum rules}

The sum rules can be easily read off from the fact that
the response function is an analytical function in the upper half
plane and falls off with large frequencies faster than 
$1/\omega^2$ such that the compact Kramers
Kronig relation reads
\be
\int d \omega ' {\chi(\omega')\over \omega'-\omega +i0}=0
\ee
closing the contour of integration in the upper half plane. From this one has
\be
{\rm Re} \chi(\omega)&=&\int {d \omega' \over \pi} {{\rm Im}
  \chi(\omega')\over \omega'-\omega}
\nonumber\\
&=&{<\omega>\over \omega^2}+{<\omega^3>\over \omega^4}+...
\label{defm}
\ee
with the moments
\be
<\omega^{2 k+1}>=\int {d \omega \over \pi}  \omega^{2 k+1}{\rm Im}
  \chi(\omega).
\ee
The first two moments are known exactly to be (appendix~\ref{pert})
\be
<\omega>=\int {d \omega \over \pi}  \omega{\rm Im}
  \chi(\omega)={n q^2\over m }
\label{s}
\ee
with the density $n$ and the mass $m$ of the particles and
\cite{IKP84,PV73,P65}
\be
<\omega^3>&=&\int {d \omega \over \pi}  \omega^3 {\rm Im}
  \chi(\omega)=2 E_{int} {q^4\over m^2}+{n q^6 \over 4 m^3}
\nonumber\\&&
-{n^2 q^4
    \over m^2} V_q \tilde I(q).
\label{ex}
\ee
Here $E_{int}$ is the kinetic energy of the interacting system and
\be
\tilde I(q)&=&\!-\!{1\over n} \int {d k \over (2 \pi)^3} (S_{k-q}\!-\!S_k\!+\!n \delta_{k,q}\!-\!n\delta_{k,0})
  {(k\cdot q)^2\over q^4} {V_k\over V_q}
\nonumber\\
&=&I(q)-1
\label{I}
\ee
where $I(q)$ is usually presented in literature \cite{IKP84,PV73}
\be
&&I(q)=-{1\over n} \int {d k \over (2 \pi)^3} (S_{k-q}-S_k)
  {(k\cdot q)^2\over q^4} {V_k\over V_q}.
\ee
and $S_k$ is the structure factor (\ref{S}).

In order to understand the different
contributions, the short distance limit (\ref{l}) 
from (\ref{I}) is performed
\be
&&\lim\limits_{q\to \infty}\tilde I(q)=-{1\over n} \int {d k \over (2 \pi)^3}
(S_{k}-1 +n \delta_{k,0})
\nonumber\\&&
\qquad \qquad  \times \left [ {((k+q)\cdot q)^2\over q^4} {V_{k+q}\over V_q}- {(k\cdot q)^2\over q^4} {V_k\over V_q}\right ]
\nonumber\\&&=
(1-g_0)+\lim\limits_{q\to \infty}{1\over n} \int {d k \over (2 \pi)^3} (S_{k}-1)
{(k\cdot q)^2\over q^4} {V_k\over V_q} -1
\label{I2}
\ee
where the last term comes from the $\delta_{k,0}$ term. The first term
alone is sometimes called exact result \cite{IKP84,STLS68,S70,K73},
which holds only for static local fields. The second
term describes the motion of particles inside the correlation
hole and takes for
Coulomb just $-\frac 1 3 (1-g_0)$ which has been pointed out in
\cite{N74}.
Together one obtains the small distance result
\be
&&\lim\limits_{q\to \infty}\tilde I(q)=\frac 2 3 (1-g_0)-1
\label{q}
\ee
in agreement with the direct expansion (\ref{l}).

Now we proceed and derive the boundaries for the local field
$f_q(\omega)$ from (\ref{chim}) by the above sum rules.
Therefore we look at the large $\omega$ expansion
of (\ref{chim}) from which we can check with the help of
(\ref{defm}) the desired sum rules (\ref{s}) and (\ref{ex}). The
simple RPA leads to [see also (\ref{pnn})-(\ref{pnje})]
\be
&&\Pi_0(\omega)=
{n q^2\over m \omega^2}
+\left (
2 E {q^4\over m^2} +{n q^6 \over 4 m^3} 
\right ) 
{1\over \omega^4}+o(1/\omega^5),
\ee
from which one gets with (\ref{chim})
\be
&&\chi(\omega)=
{n q^2\over m \omega^2}
+\left (
2 E {q^4\over m^2} +{n q^6 \over 4 m^3} 
+{n^2 q^4 \over m^2}(V_q+f_q(\infty))\right ) 
{1\over \omega^4}
\nonumber\\&&+o(1/\omega^5).
\label{sum3}
\ee
The first order energy weighted sum rule (\ref{s}) is fulfilled
trivially 
provided $\Pi$ fulfills it. The third order sum rule
(\ref{ex}) can be fulfilled if we construct the
local field according to \cite{IKP84}
\be
f_q(\infty)&=&-V_q (1+\tilde I(q))-{2 \over n^2} (E-E_{int})
\nonumber\\
&=&\!-\!V_q I(q)-{2 \over n^2} (E\!-\!E_{int})\label{req}
\ee
where $E$ is the kinetic energy of the noninteracting
system. The last term describes the fact that the third order
frequency sum rule of the polarization function yields the
noninteracting kinetic energy. This form neglecting the last term has
been discussed in \cite{PV73}. In the later derivation of the
polarization function we cannot consider the kinetic energy anymore as
interaction free, since the relaxation time appears as well as the
effective mass. Therefore within the quasiparticle picture used here the
difference $E-E_{int}$ vanishes or positively stated, is accounted for
by the effective mass. To facilitate the comparison with 
the literature we have kept this
difference formal as $\Delta E=-{2\over n^2} (E-E_{int})$. 

With (\ref{req}) we have given the constraint on the dynamical local
field from the third order frequency sum rule. In the following we
will present a dynamical local field which fulfills both
requirements, the compressibility (\ref{fexp}) and the frequency sum
rule (\ref{req}).

\section{Dynamical response function}

In \cite{MF99,MF00}
was given the polarization function for an interacting quantum system 
imposing conservation laws on the relaxation time approximation.  
These polarization functions we have denoted by $\Pi^{\rm n}$ for density
conservation imposed, $\Pi^{\rm n,j}$ for density and current
conservation and $\Pi^{\rm n,j,E}$ for density, current and energy
conservation. In the former paper we could give only formal matrix 
expressions for the
response function. In appendix~\ref{dynres} we repeat shortly the way
of derivation from the quantum kinetic theory and give now the explicit
form  of the response function. We obtain with (\ref{chia}) and (\ref{chi1a})
\be
\chi(\omega)={\Pi^{\rm n,j,E}(\omega)\over 1-V_0 \Pi^{\rm n,j,E}(\omega) -2 m V_4 \Pi_{13}(\omega)}
\label{chi}
\ee
where
\be
\Pi_{13}(\omega)&=&{\Pi^{\rm n,j,E}\over 2 m}\,\,\,\,
{
\Pi_h \Pi_2(0)-i \tau \omega {\Pi_2(\omega+{i\over \tau})}
\over 
\Pi_h \Pi_0(0)-i \tau \omega {\Pi_0(\omega+{i\over \tau})}
}
\nonumber\\
\Pi_h&=&
{\Pi_2^2(\omega+{i\over \tau})-\Pi_0(\omega+{i\over \tau}) \Pi_4(\omega+{i\over \tau}) \over\Pi_2^2(0)-\Pi_0(0) \Pi_4(0)}
\label{chi1}
\ee
and
\be
V_0&=&{\delta \over \delta n} \Delta-V_4 {q^2\over 4}+V_q \nonumber\\
V_4&=&{\delta \over \delta n} {1\over 2 m}=-{1\over 2 n m} {\partial
  \ln m\over \partial \ln n}.
\label{V04}
\ee
The explicit expressions of the moments of the polarization function
are summarized in appendix~\ref{a} for computation.
From the response function (\ref{chi}) we will read off the main
result of 
this paper: the dynamical local field for quasiparticles with
the effective mass and energy. Now we are going to
work out explicitly the form of local field to 
show that the third order sum rule can be fulfilled and the correct
compressibility is obtained.

\subsection{Dynamical local field}

Comparing (\ref{chi}) with (\ref{chim}) an intermediate dynamical local field can
be read off as
[$V_0={\delta\over
  \delta n} \Delta- V_4 q^2/4+V_q$]
\be
\tilde f_q(\omega)&=&{\delta\over
  \delta n} \Delta \!+\!V_4 \left ({
\Pi_h \Pi_2(0)-i \tau \omega {\Pi_2(\omega+{i\over \tau})}
\over 
\Pi_h \Pi_0(0)-i \tau \omega {\Pi_0(\omega+{i\over \tau})}
}\!-\!{q^2\over 4} \right )
\nonumber\\
&=&{\delta\over
  \delta n} \Delta \!+\!V_4 
\left ({
\Pi_2(0) \over  \Pi_0(0) }\!-\!{q^2\over 4} 
\right )+o({1\over \omega})
\nonumber\\
&=&{\delta \over \delta n} \Delta-{\partial \ln m \over \partial
  \ln n}
\left \{ \matrix { {3  \over 2 n^2 K_0}+o(q^2)
\cr
{1  \over 3} {E   \over   n^2}+o(1/q^2)
}+o({1\over \omega})
\right .
\nonumber\\&&
\label{exp1}
\ee
where we used the expansion of appendix~\ref{a}, explicitly 
(\ref{b16}), 
in the first line and (\ref{ex0}) and (\ref{ex2}) in the last
line. Please remark that $\Pi^{\rm n,j,E}$ itself contains correlations
beyond the polarization in RPA, $\Pi_0$. This we will present in a moment.

First, we see the astonishing result that obviously
\be
\lim\limits_{\omega \to 0}\tilde f_q(\omega)=
\lim\limits_{\omega \to \infty}\tilde f_q(\omega)
\label{eston}
\ee
the static local field required for the compressibility 
agrees with the infinite frequency limit
required for the third order sum rule (\ref{sum3}).
\footnote{One should not be mislead to
  the conclusion that this violates the Kramers Kronig relation for
  $\tilde f_q(\omega)$. A toy example of ${\rm Im} \tilde f=\sin{(a
    \omega)}/(\omega^2-4\pi^2/a^2)$ and the corresponding real part obtained from
  the Kramers Kronig relation (\ref{defm}) 
shows that indeed (\ref{eston}) can hold simultaneously with the
Kramers Kronig relation.} 
This shows that
the answer to the sumrule puzzle is not due to the explicit 
dynamical character of the local field
as often claimed in the literature. Instead we will see in
paragraph~\ref{sumr} 
that it is due to the underlying selfenergy correction which
has to obey certain relations.

At this point it is important to avoid a
misunderstanding. The intermediate dynamical local field, $\tilde
f_q$, is not the total
one describing correlations beyond RPA polarization function,
$\Pi_0$, which would be the case only in the infinite frequency limit. 
Instead, part of the correlations are already captured in
$\Pi^{\rm n,j,E}$. To make this explicit we write (\ref{chi}) as
\be
{1\over \chi(\omega)}&=&{1\over \Pi^{\rm
    n,j,E}(\omega)}-V_q-\tilde f_q(\omega)
\nonumber\\&=&{1\over \Pi_0(\omega)}-V_q-\tilde f_q(\omega)+\tilde f_q^*(\omega)
\ee
where the difference between $\Pi^{\rm n,j,E}$
and $\Pi_0$ has been recasted into a local field contribution
\cite{MF00} derived from (\ref{result})-(\ref{M-n}),
\be
\tilde f_q^*(\omega)&=&{1 \over \Pi^{\rm n,j,E}(\omega)}-{1 \over \Pi_{0}(\omega)}\nonumber\\
&=&-{1\over
  1-i\omega \tau} \left ({1\over \partial_\mu n}-{2 E\over n^2}\right
)+o(q^2)
\nonumber\\
&=&{1\over
  1-i\omega \tau} {8 \epsilon_f \over 15 n} +o(q^4)
\label{resul}
\ee
with the last line valid for zero temperature. Oppositely in the
static limit $\tilde f_q^*(0)=0$. Together with
(\ref{exp1}) we obtain an effective local field renormalizing the RPA
\be
f_q^{\rm eff}(\omega)&=&\tilde f_q(\omega)-\tilde f_q^*(\omega)\nonumber\\
&=&{\delta \over \delta n} \Delta-{\partial \ln m \over \partial
  \ln n}
 {3  \over 2 n^2 K_0}+{{1\over \partial_\mu n}-{2 E\over n^2}\over
  1-i\omega \tau} +o(q^2)\nonumber\\
&=&{\delta \over \delta n} \Delta-{\epsilon_f \over n}\left (
{\partial \ln m \over \partial
  \ln n}+
{1\over
  1-i\omega \tau} \frac {8}{ 15} \right )
\nonumber\\&&
\label{local}
\ee
where the last line is again the zero temperature limit. The high
frequency limit required for the third order sum rule all agree
$f_q^{\rm eff}(\infty)=\tilde f_q(\infty)=f_q(\infty)$.

\subsection{Connection to density functionals}

In order to establish the connection to the ground state exact
relations \cite{GK85} of exchange correlation energy, $\varepsilon_{\rm
  xc}$, we see from (\ref{local})
\be
&&\lim\limits_{\omega\to 0} \lim\limits_{q\to 0} f^{\rm eff}_q(\omega)\!\equiv\! {d^2\over d n^2}[n
\varepsilon_{\rm xc}(n)]\!=\!{\delta \over \delta n} \!\Delta\!-\!{\epsilon_f \over n}\!\left (\!
{\partial \ln m \over \partial
  \ln n}\!+\!
\frac {8}{ 15}\! \right )
\nonumber\\&&
\ee
and
\be
&&\lim\limits_{q\to 0} f^{\rm eff}_q(\infty)\equiv - \frac 4 5 n^{2/3}
{d\over d n}\left [
{\varepsilon_{\rm xc}(n)\over n^{2/3}}\right ]={\delta \over \delta n} \Delta-{\epsilon_f \over n}
{\partial \ln m \over \partial
  \ln n}
\nonumber\\&&
\ee
with $f^{\rm eff}_0(0)<f^{\rm eff}_0(\infty)<0$. This establishes the
link to time dependent density functional theories
\cite{D86,FHER93,NCT98}. A overview about different approximation
schemes are given in \cite{I82}. We will see in paragraph~\ref{sumr}
that in order to fulfill the third order frequency sum rule we have to
have an additional effective mass beyond $\varepsilon_{\rm xc}$.

\subsection{Compressibility sum rule}

From (\ref{exp1}) one sees immediately that the required form for the
compressibility sum rule (\ref{fexp}) appears, since $\Pi^{\rm n,j,E}(0)=\Pi_0(0)$.
Therefore we have derived a dynamical response
function and a local field which shows in the static limit the correct
compressibility (\ref{direct}).
This 
compressibility formulae (\ref{direct})  
can be checked alternatively by calculating explicitly the frequency
integral in (\ref{compress}). The required small wavevector limit of the
so called screened structure function (\ref{compress}) takes the form
[$\Pi=\Pi^{\rm n,j,E}$]
\be
&&{n\over \beta} K
=\lim\limits_{q\to 0}{1\over n \pi}
\int {d \omega \over 1-{\rm e}^{-\beta \omega}} 
\nonumber\\
&&\times
{{\rm Im}\Pi(\omega)(1-{\bar f_q}(\omega) {\rm Re}\Pi(\omega))+ {\rm Im}\Pi(\omega) {\bar f_q}(\omega) {\rm
    Re}\Pi(\omega)\over (1-{\bar f_q}(\omega) {\rm
    Re}\Pi(\omega))^2+({\bar f_q}(\omega) {\rm Im}\Pi(\omega))^2}
\nonumber\\&&
\label{w1}
\ee
where $\tilde f_q(\omega)={\bar f_q}(\omega)-V_q$.
Now we observe that
\be
&&\lim\limits_{q\to 0}{\rm Im}\Pi(\omega)=\lim\limits_{q\to 0}{\rm
  Im}\Pi_0(\omega)
\nonumber\\&&
=\!-\!\pi \int {dp \over (2 \pi )^3} ({\rm e}^{-\beta
  \omega}\!-\!1)\delta(\omega\!-\!{p\cdot q\over m}) F(p)(1\!-\!F(p))\!+\!o(q^2)
\label{w11}\nonumber\\&&
=0+o(q^2)
\ee
vanishes for small $q$. Therefore we have to perform the limit
in (\ref{w1}) in the distribution sense to obtain
\be
{n\over \beta} K&=&\lim\limits_{q\to 0}{1\over n \pi}
\int {d \omega \over 1-{\rm e}^{-\beta \omega}}
\nonumber\\&
\times&\left ({{\rm Im}\Pi_0(\omega) \over  1-{\bar f_q} {\rm Re}\Pi(\omega)}+\pi {\rm Re}\Pi(\omega)  \delta
  (1-{\bar f_q} {\rm Re}\Pi(\omega)) \right ).
\nonumber\\&&\label{w2}
\ee
It is not difficult to see that the second part vanishes and
we obtain
\be
{n\over \beta} K&=&{1 \over n} {\int
{d p \over (2 \pi)^3} F(p)(1-F(p)) 
\over 1-{\bar f_q} {\rm Re} \Pi(0)}
\nonumber\\
&=&{1\over n \beta} {n^2 K_0\over 1+n^2 K_0 {\partial \Delta \over
    \partial n}-\frac 3 2 {\partial
  \ln m\over \partial \ln n}}
\ee
which agrees with (\ref{direct}).

Therefore, we have shown that the dynamical local field
(\ref{exp1}) from the response function (\ref{chi}) leads  to the same
compressibility  (\ref{compress}). This gives besides the
compressibility sum rule already checked a second proof that we
have derived a dynamical local field which leads to the correct
compressibility. The static limit will allow now to
complete compressibility and third order frequency sum rule simultaneously.

\subsection{Frequency sum rules}\label{sumr}

Now that we have the response function
(\ref{chi}) at hand  we can proceed
and proof the frequency sum rules (\ref{s}) and (\ref{ex}) explicitly. 
First we expand the polarization functions for large
frequencies. The superscript denotes which conservation laws are
obeyed, density $(n)$, energy $(E)$ and current $(j)$ respectively. We
obtain
\be
&&\Pi^{\rm n}(\omega)=
{n q^2\over m \omega^2}-i{n q^2\over m \omega^3  \tau}
+\left (
2 E {q^4\over m^2} +{n q^6 \over 4 m^3}-{n q^2\over m \tau^2} 
\right )  {1\over \omega^4}
\nonumber\\
&&-i {q^4\over m^2\omega^5 \tau} 
\left ({n^2 \over \Pi_0(0)}+6 E -{n m\over q^2 \tau^2}+ {3 n q^2 \over
  4 m}\right )+o\left ({1\over \omega^6}\right )
\nonumber\\&&
\label{pnn}\\
&&\Pi^{\rm n,E}(\omega)=\Pi^{\rm n}(\omega)+o\left ({1\over \omega^5}\right )
\label{pne}
\\
&&\Pi^{\rm n,j}(\omega)=
{n q^2\over m \omega^2}
+\left (
2 E {q^4\over m^2} +{n q^6 \over 4 m^3} 
\right ) 
{1\over \omega^4}
\nonumber\\
&&-
i {q^4\over m^2\omega^5 \tau} 
\left ({n^2 \over \Pi_0(0)}+2 E + { n q^2 \over
  4 m}\right )+o\left ({1\over \omega^6}\right )
\label{pnj}\\
&&\Pi^{\rm n,j,E}(\omega)=\Pi^{\rm n,j}(\omega)+o\left ({1\over \omega^5}\right ).
\label{pnje}
\ee
We see that the current conservation repairs some defiances of the
Mermin-Das polarization function, $\Pi^{\rm n}$ which obeys only density conservation,
in that the imaginary part shows a
different frequency behavior 
\be
\lim_{\omega\to\infty}{\rm Im}\Pi^{\rm n,j,E}({q},\omega)&=&
\frac{n^2 q^4}{\omega^5 \tau m^2}\left ({1\over \partial_\mu n}-{2 E\over n^2}\right
)\nonumber\\
\label{njwim}\\
\lim_{\omega\to\infty}{\rm Im}\Pi^{\rm n}({q},\omega)&=&
-\frac{n q^2}{\omega^3 \tau m}.
\label{nwim} 
\ee 
The last formulae corrects a misprint in formula (23) of 
\cite{MF00}. This different behavior of the imaginary part is also
reflected in different expressions for the third order moment
$(o(1/\omega^4))$ or third order sum
rule.

From (\ref{defm}) and (\ref{pnn})-(\ref{pnje}) we read off
the sum rules
\be
<\omega>=\int {d \omega \over \pi}  \omega \, {\rm Im}
  \Pi(\omega)={n q^2\over m }
\ee
which holds for each $\Pi^{\rm n}$, $\Pi^{\rm n,j}$, $\Pi^{\rm n,j,E}$
and $\chi$.
In contrast to that we will see now that the third order sum rule gives different
results for the inclusion of different conservations laws. Using the
polarization function including density, energy and momentum
conservation we obtain from (\ref{pnj})
\be
<\omega^3>=\int {d \omega \over \pi}  \omega^3 {\rm Im}
  \Pi^{\rm n,j,E}(\omega)=2 E {q^4\over m^2}+{n q^6 \over 4 m^3}.
\label{w30}
\ee
We remark that according to (\ref{pnn})-(\ref{pnje}) the Mermin-Das polarization (\ref{M-n}) including only density
conservation or even including additionally energy conservation
(\ref{M-ne}) 
would yield an additional $-n q^2/m \tau^2$ term which is
an artefact. This is repaired by additionally taking into account
momentum conservation.

Comparing (\ref{ex}) with (\ref{w30}) we see that just the last terms
are missing. In order to obtain this sum
rule we have to use the response function (\ref{chi}) and not the
polarization function
for which this sum rule is actually designed. With (\ref{exp1}) and
(\ref{sum3})
one gets
\be
&&\chi(\omega)=\Pi^{\rm n,j,E}(\omega)\!+\!{n^2 q^4\over m^2} \left (
  V_0\!+\!{\Pi_2(0)\over \Pi_0(0)} V_4 \right ) {1\over \omega^4}\!+\!o\left
    ({1\over \omega^5}\right ).\nonumber\\
&&
\ee
Consequently, the third order sum rule (\ref{req}) is rendered correctly
if one sets
\be
f_q(\infty)&=&V_0\!-\!V_q\!+\!V_4 {\Pi_2(0)\over \Pi_0(0)}={\delta \Delta \over
  \delta n} \!+\!V_4 \left ({\Pi_2(0)\over \Pi_0(0)}\!-\!{q^2\over 4}\right )
\nonumber\\
&\equiv& -V_q(1+\tilde I(q))+\Delta E.
\label{req1}
\ee
By the
requirement (\ref{req1}) we have a possibility to fulfill the third
order sumrule exactly from the dynamical response as well as static
local field model.

\subsection{Consequences on selfenergies}

Let us now work out what that means for our
selfenergy parameterization $\Delta$ and $m$. From (\ref{req1}) 
we obtain a determining condition for the 
effective mass and energy shift [$I(q)=1+\tilde I(q)$]
\be
&&V_q I(q)
={1\over 2 m n}\left ( {\Pi_2(0)\over \Pi_0(0)}-{q^2\over 4} \right ) {\partial
  \ln m\over \partial \ln n}-{\delta \Delta \over
  \delta n}+\Delta E.
\nonumber\\&&
\label{mass}
\ee
Since we work with the effective mass and shift parameterization of the
quasiparticle energy the difference between $E$ and $E_{int}$, $\Delta
E$,
vanishes but we keep it for completeness further on.  

Applying the small wavevector limit (\ref{small}) and (\ref{exp1}) 
 we see now from (\ref{mass}) for
Coulomb systems 
\be
\lim\limits_{q\to0}\left ({\delta \Delta \over \delta n}-\frac 3 2 {1\over n^2 K_0} {\partial
  \ln m\over \partial \ln n}\right )
&=&\Delta E -\frac 2 5 {q^2 V_q \over k_f^2 }\gamma+o(q^2).
\nonumber\\&&
\label{d1}
\ee
Oppositely from (\ref{mass}), the large wavevector or small distance limit
(\ref{l}) and (\ref{exp1})reads
\be
\lim\limits_{q\to\infty}\left ({\delta \Delta \over \delta n}-\frac 1 3 {E\over n^2} {\partial
  \ln m\over \partial \ln n}\right )&=&
\Delta E -\frac 2 3 (1-g_0) V_q +o(1/q^2)
\nonumber\\&&
\label{d2}
\ee
where the last term on the right side vanishes for Coulomb potentials
and persists only for potentials which range falls faster than Coulomb. 

If we assume homogeneous systems, where $\Delta$ and $m$ becomes
independent of the wavevector, the equations (\ref{d1}) and (\ref{d2})
determines the quasiparticle shift as well as the effective mass via
\be
{\partial
  \ln m\over \partial \ln n}&=&{{2 q^2 V_q \over 5 k_f^2} \gamma
 \over \frac 3 2
  {1\over n^2 K_0}-\frac 1 3 {E\over n^2}}\nonumber\\
{\delta \Delta \over \delta n}&=&{{2 q^2 V_q \over 5 k_f^2} \gamma
  \over \frac 9 2
  {1\over E K_0}-1}+\Delta E  -\frac 2 3 (1-g_0) V_q +o(1/q^2)
\label{mn}
\ee
where the last term vanishes for Coulomb potentials.
If we remember the relation between the selfenergy $\sigma$ and the
effective mass (\ref{mass1}) we can determine from (\ref{mn}) the
thermal averaged selfenergy and wave function renormalization
${\cal Z}=(1-\partial_\omega \sigma)^{-1}$. For the special case of zero
temperature 
and
neglecting the usually small $\partial_p \sigma$ part, it
reads
\be
\Delta=\sigma|_{\epsilon=\epsilon_f}&=&{1\over 10} q^2 V_q \int\limits^n_0 dn'
{\gamma(n')\over p_f^2(n')}+\int\limits_0^n d n' \Delta E\nonumber\\
\ln{{\cal Z}}|_{\epsilon=\epsilon_f}&=&
- m q^2 V_q \int\limits^n_0 dn'
{\gamma(n')\over p_f^4(n')}
\ee
with $\gamma$ from (\ref{gamma}) such that the effective mass takes the form
\be
m=m_0 \, \exp{\left \{ m q^2 V_q \int\limits^n_0 dn'
{\gamma(n')\over p_f^4(n')}\right \}}.
\ee
The general expression for finite temperatures is given by (\ref{mn}).

Since the needed expressions $\tilde I$ in (\ref{I}) or $\gamma$ in
(\ref{gamma}) are functions of the structure function which is given 
itself again by the response function (\ref{S}) we
have the usual self-consistent procedure analogous
to (\ref{STLS}) first introduced by Singwi Sj\"olander \cite{STLS68} but
with another $G(q)$. Here we  suggest to obtain the effective mass
$m$ and energy shift by (\ref{mass}) using $I(q)$.

Alternatively one
might use the well known experimental values of the static structure
factor $S_q$ and determine by this way the effective mass and energy
shift. 
This definition of effective mass has the advantage that the third order
and compressibility sum rule of the response function will be rendered exactly.
Therefore equation (\ref{mn}) is the second main
result of this paper.

Let us remark that if we would have no effective mass but a mere
density dependent selfenergy $\sigma$ like in density functional theories, the requirement of (\ref{d1}) and (\ref{d2})
corresponding to the compressibility and third order frequency sum
rule cannot be fulfilled simultaneously. This was remarked in detail in
literature \cite{VG73,I84}. By including the effective mass we can resolve
this puzzle here.

\section{Numerical results and discussion}

As a test example we will now consider the unpolarized electron gas in
three dimensions (3D) at zero
temperature. The density parameter
is the usual Bruckner parameter, defined as the ratio of interparticle distance
to the Bohr radius, $r_s= (3/4 \pi n)^{1/3}/a_0$.
First we will give a simple quasiparticle picture (QP), where the $\Delta$
and the effective mass are determined as density--dependent constants
from MC data. In the second step, we will allow  that the $\Delta$ and
the effective mass depend on the wavevector. This  will lead to the
self-consistent quasiparticle picture (SQP).

In order to calculate the quasiparticle parameter $\Delta$ and
the effective mass, we employ the results of Monte Carlo (MC) simulations
\cite{OB94,MCS95}. For a parameterization of the MC data see
\cite{CSOP98}. 
In \cite{IKP84} was discussed the difference
of interacting and free kinetic energies, $\Delta E={2\over
  n^2}(E_{int}-E)$. This difference is given as
$\delta=E_{int}/E-1$ by the MC data
of Ceperley and Alder. In our quasiparticle picture
one has $E_{int}/E=m_0/m$, and the effective mass is given by
$m=m_0/(1+\delta)$. This allows us to determine the needed derivative as
\be
{\partial \ln m\over \partial \ln n}={r_s\over 3(1+\delta)} {\partial
  \delta \over \partial r_s}.
\label{mas}
\ee 

\parbox[h]{8cm}{
\begin{table}[h]
\begin{tabular}[]{|l||c|c|c||c|c|c|}
\hline
$r_s$& $\gamma_0$& $\delta$ &\multicolumn{4}{c|}{$1-g(0)$}\\
& MC & MC & MC & SQP &QP &PV+$\Delta$E 
\\
\hline
1 & 0.2567 &0.036  & 0.7276 &0.7240   &  0.6626  &0.6971\\
2 &        &0.091  &(0.8627)&0.8627   &  0.7666  &0.8346\\
3 & 0.2722 &       & 0.9078 &0.9201   &  0.8456  &\\
5 & 0.2850 &0.292  & 0.9768 &0.9627   &  0.9351  &1.0558\\
10& 0.3079 &0.619  & 0.9976 &0.9733   &  0.9613  &1.1231\\
\hline
\end{tabular}
\caption{The long wavelength limit of the local field factor,
  $\gamma_0$, and the small range value of the pair correlation according
  to the MC data of \protect\cite{OB94}. The value in brackets is an interpolation. The quasiparticle (QP),
  self-consistent quasiparticle (SQP) as well as
  Pathak-Vashishta value together with $\Delta E$ (PV+$\Delta E$) is
  given for comparison. }\label{tab1}
\end{table}
}

The difference $\Delta E$ is given by the large--wavelength limit of
the local field
\be
\lim\limits_{q\to 0} G(q)&=&\gamma_0 \left ( {q\over k_f}\right
)^2
=\lim\limits_{q\to 0}{1\over V_q}(V_q I(q)- \Delta E) \nonumber\\
&=&\lim\limits_{q\to 0}{1\over V_q} \left ({1\over 2 m n} {\partial
  \ln m\over \partial \ln n}-{\delta \Delta \over
  \delta n}\right ),
\label{mas1}
\ee
which is presented by $\gamma_0$ \cite{OB94}, see
table~\ref{tab1}. This allows us to
determine the quasiparticle energy ${\delta \Delta \over
  \delta n}$, since the last line of (\ref{mas1}) is just (\ref{mass}).
The first model can be called improved Pathak/Vashishta scheme
(PV+$\Delta E$) while the second one together with the effective mass
(\ref{mas}) establishes the quasiparticle picture proposed
here. We remark that, in order to realize a certain Bruckner parameter
$r_s$,
the quasiparticle picture must be calculated with $r_s
(1+\delta)$, since all formulae work with effective mass leading to $r_s/(1+\delta)$.

\begin{figure}[h]
\psfig{file=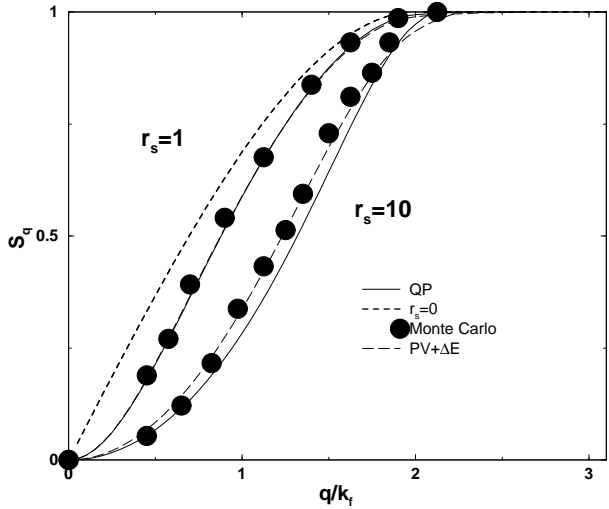,height=8cm,angle=-90}
\caption{The static structure factor for $r_s=1$ and $r_s=10$ 
from the (PV)-, the (PV+$\Delta
  E$)- and (QP)- model compared to the MC data \protect\cite{OB94} as
  presented in \protect\cite{G97}.}
\label{figs}
\end{figure}

\begin{figure}[h]
\psfig{file=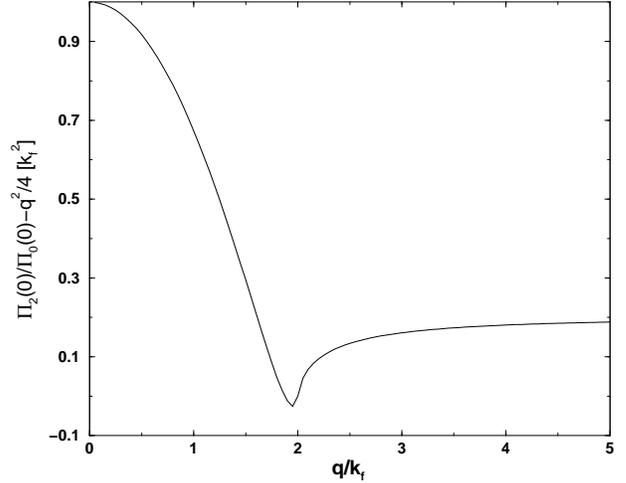,height=8cm,angle=-90}
\caption{${\Pi_2(0)\over \Pi_0(0)}-{q^2\over 4}$ of
  (\protect\ref{mass}) or explicitly (\protect\ref{massK}) for
  three dimensions at zero temperature.}
\label{help}
\end{figure}

With this determination of the QP parameters as well as (PV+$\Delta
E$) from MC data we can now compare the results of the iteration
scheme $f_q\to S_q\to I(q)\to f_q$. For zero temperature all formulae
simplifies and we give them here for convenience in dimensionless
momentum $K=q/k_f$, energy $\Omega=\omega/\epsilon_f$ and distance 
$R=r k_f/\hbar$.  
Then the dimensionless effective local field is given by
\be
f_K&=&-\left ({2 s^4
\over 9 \pi^4}\right )^{1/3} r_s \left ( {I(K)\over K^2}+\gamma_0-\frac 2 5
  \gamma \right )
\nonumber\\
&&-{r_s\over 4(1+\delta)}{\partial \delta \over \partial
    r_s}
\left (1-{\Pi_2(0,K)\over \Pi_0(0,K)}+{K^2\over 4} \right )
\label{fqt0}
\ee
with $\gamma_0$ from (\ref{mas1}) and $\gamma=-\frac 1 s
\int\limits_0^\infty d K (S_K-1)$ from (\ref{gamma}) and the spin
degeneracy for electrons $s=2$. The (PV) model
would consists only in the term $I(K)$ on the right hand side of
(\ref{fqt0}) and the (PV+$\Delta
  E$) model takes into account the first line of (\ref{fqt0}). The
  (QP) picture finally takes all terms into account where we have used
  (\ref{req1}) or (\ref{mass}) and the requirement (\ref{mas1}). 
The dimensionless function, ${\Pi_2(0,K)\over
  \Pi_0(0,K)}={\Pi_2(0)\over k_f^2 \Pi_0(0)}$ plotted in figure~\ref{help}, is given by
\be
&&{\Pi_2(0,K)\over \Pi_0(0,K)}={1\over 2} {3-{K^2\over 4}-{1\over 4 K} \left
    (1-{K^2\over 4}\right )^2 \ln{|{2-K\over 2+K}|}\over
1-{1\over 4 K} \left
    (1-{K^2\over 4}\right ) \ln{|{2-K\over 2+K}|}}\nonumber\\
&&={1\over 4} \left \{ 
\matrix{1-{K^2\over 3}+o(K^{4})\cr
{K^2\over 4}+\frac 1 5 -{48\over 175 K^2}+o(K^{-4})
}
\label{massK}
\right ..
\ee

Provided we know the effective local field, $f_K$ in (\ref{fqt0}), the static structure
factor can be obtained from (\ref{S}) as
\be
S_K={3 \over 4 \pi} \int\limits_0^\infty d \Omega \, {\rm Im }\, 
{\Pi_0(\Omega, K)\over 1-\left ( f_K+({2 s^4
\over 9 \pi^4})^{1/3} {r_s \over K^2}\right
  ) \Pi_0(\Omega,K)}
\ee
where we have used the zero temperature dimensionless quantum polarization
from (\ref{pn})
\be
&&\Pi_0(\Omega,K)=-1+{4 K^2-(\Omega-K^2)^2\over 8 K^3} \ln {2
  K+\Omega-K^2\over 2 K-\Omega+K^2}
\nonumber\\&&\qquad\qquad\quad\quad\quad
-{4 K^2-(\Omega+K^2)^2\over 8 K^3} \ln {2
  K+\Omega+K^2\over 2 K-\Omega-K^2}\nonumber\\
&&+i \frac \pi 2 \left \{ 
\begin{array}{ll}
0&\\
{\Omega\over K} &{\rm for}\, K<2\, {\rm and}\, |\Omega|<|K^2-2 K|\\
{4 K^2-(\Omega-K^2)^2\over 4 K^3}&{\rm for}\,|K^2-2 K|<|\Omega|<|K^2+2 K|
\end{array}
\right ..
\nonumber\\&&
\ee
With the help of the static structure factor we have the
pair correlation
function (\ref{gs}) 
\be
g_R=1+{3\over s R} \int \limits_0^\infty d K \, K\, \sin{(K R)} \left (
  S_K-1\right )
\ee
from which
the required
$I(K)$ function reads according to (\ref{cpot})
\be
I(K)&=&-2 \int \limits_0^{\infty} {d R\over R} \left (g_R-1\right ) j_2(K R)
\ee
with the spherical Bessel function $j_2(x)$. This function now enters
(\ref{fqt0}) closing the iteration. 

\begin{figure}[h]
\psfig{file=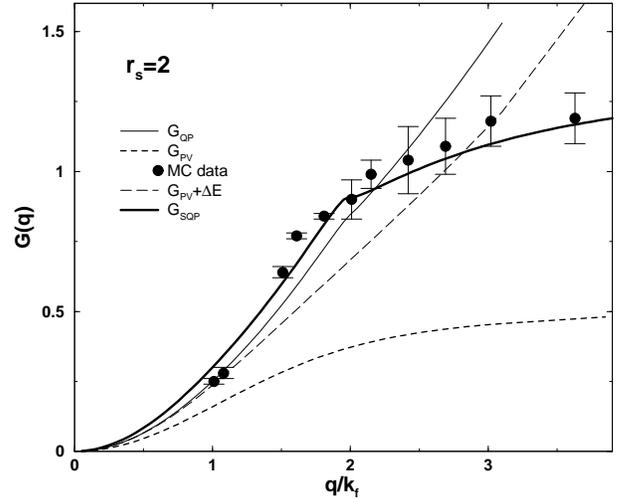,height=8cm,angle=-90}
\caption{The static local field for $r_s=2$  
from the (PV)-, the (PV+$\Delta
  E$)- and (QP)- model compared to the MC data \protect\cite{MCS95}.}
\label{figg2}
\end{figure}

In table~\ref{tab1} we compare the small distance value of the
pair correlation (\ref{gs}) of the (PV+$\Delta E$) model with the (QP)
model. We see that for more dense systems the (PV+$\Delta E$) model
leads to correlations which are too large while the (QP) model is lower than the MC
values at higher $r_s$. In figure \ref{figs} we compare the static structure factor of the 
two models with the MC data. We see that a difference occurs
between the (PV+$\Delta E$) and (QP) model at higher $r_s$.

\begin{figure}[h]
\psfig{file=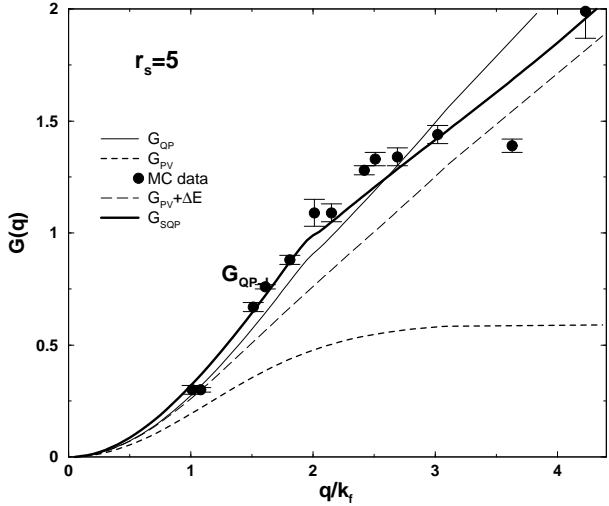,height=8cm,angle=-90}
\caption{The static local field for $r_s=5$  
from the (PV)-, the (PV+$\Delta
  E$)- and (QP)- model compared to the MC data \protect\cite{MCS95}.}
\label{figg5}
\end{figure}

\begin{figure}[h]
\psfig{file=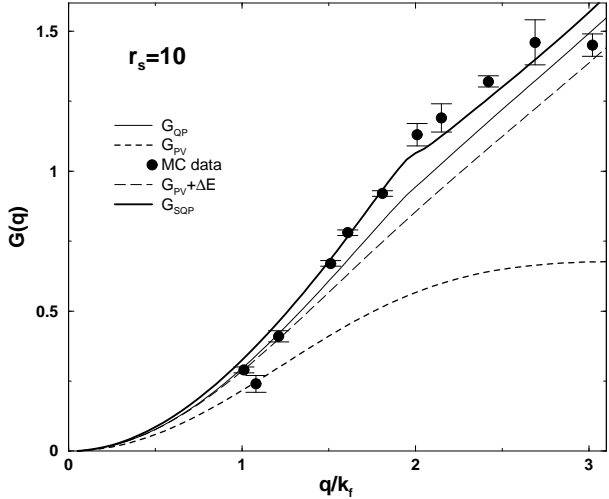,height=8cm,angle=-90}
\caption{The static local field for $r_s=10$  
from the (PV)-, the (PV+$\Delta
  E$)- and (QP)- model compared to the MC data \protect\cite{MCS95}.}
\label{figg10}
\end{figure}

The difference between the models becomes more apparent if we plot the
local field factor as in figures \ref{figg2}-\ref{figg10}.
We see that the simple (PV) model underestimates the MC data though it
satisfies the third order frequency sum rule. This result is improved
by adding the $\Delta E$ read off from the large wavelength limit of MC
data. Further improvement is achieved in the (QP)
picture. At smaller densities, $r_s=2$, the local field is
overestimated at higher wavevectors which leads to the deviation seen
in table~\ref{tab1}.

The (QP) curves
show a small hump at $q=2 k_f$ in contrast to the (PV+$\Delta
  E$) model. This comes from the function ${\Pi_2(0)\over \Pi_0(0)}$
which is plotted in figure \ref{help}.
While in 3D this hump at $q=2 k_f$ is not much pronounced, it shows up in 2D
systems \cite{DPGT01}. With the help of the formulae in appendix
\ref{a} one can present also the 2D results which should be devoted to
another paper.

According to (\ref{mass}) the functional form of $\Pi_2/\Pi_0$ on
the right hand side should be equal to $I(k)$ which is a smooth function
according to figures \ref{figg2}-\ref{figg10}. Therefore with the 
simple quasiparticle picture
considered so far one cannot satisfy the third order sum rule for all $q$. In order to achieve this,
we must allow $\Delta$ as well as the effective mass to carry a
$q$-dependence. Of course this leads to a self-consistent quasiparticle
picture, since the energies under integration of the polarization
function now change their dispersion. The iteration scheme is therefore
enlarged to $f_q\to S_q\to I(q)\to\Delta(q),m(q)\to f_q$ according to
(\ref{mass}). We call this iteration here self-consistent quasiparticle
picture (SQP). It satisfies the third order frequency sum rule and the
compressibility sum rule simultaneously for all $q$. The results
describe the MC data in figures
\ref{figg2}-\ref{figg10}. Also, the small distance value of the pair
correlation function is now in better agreement with the MC data as
can be seen in table \ref{tab1}.

\section{Summary}

We have derived a response function in the quasiparticle
picture where the correlations are parameterized by a density--dependent
effective mass, energy shift and a relaxation time respecting density,
energy and momentum conservation. The
dynamical response function can be given in the form of a modified RPA including a
dynamical local field. This local field leads in the static limit to
the correct compressibility. The effective mass and quasiparticle energy
shift are proposed to be determined
by the requirement of the third order sum rule. This allows in turn
to satisfy the compressibility sum rule simultaneously. Since the
effective mass is now a function of the structure factor, one might use
experimentally known values or simulation results 
to construct a more realistic
quasiparticle picture. The consequences on microscopic expressions of
the selfenergy are presented. The thermal averaged selfenergy and
wave function renormalization, i.e. the frequency derivative of the
selfenergy, are linked to the pair correlation function at small distances. 

While all the expressions we derived are valid for finite temperatures, we
have compared them as a test example with the Monte Carlo data for an electron
gas at zero temperature. We find an improved description by the
Pathak-Vashishta scheme accomplished by an energy shift derived from
MC data. The best agreement with the data is achieved by constructing
a nonlocal quasiparticle picture allowing for a wavevector--dependent
effective mass and an energy shift in the form of a self-consistent
quasiparticle picture.

\acknowledgements
The fruitful discussions with Peter Fulde and Paul Ziesche are
gratefully acknowledged. To Marco Ameduri I am indebted for
a critical reading.

\appendix

\section{Dynamic Response function}\label{dynres}

In the former paper \cite{MF99,MF00}
it was derived formally the density, current and energy response $\{{\chi,
\chi_J, \chi_E}\}={\cal X} \{1,{0},0\}^T$ of an interacting quantum system 
\be
&&\left (\matrix {\delta n\cr q \cdot {\delta J}\cr \delta E} \right
)=\left (\matrix{\chi \cr { \chi_J} \cr \chi_E}\right ) \,\,V^{\rm
  ext}\equiv{\cal X} \left (\matrix{1\cr {0}\cr 0}\right ) V^{\rm ext}
\label{def}
\ee
to the external perturbation $V^{\rm ext}$ provided the density,
momentum and energy are conserved. This has been achieved by
linearizing the kinetic equation for the one-particle density operator
$\hat\rho$ in relaxation time approximation
\be
\dot {\hat \rho}+i[\hat {\cal E}+\hat V^{\rm ext},\hat \rho] ={\hat \rho^{\rm l.e.}-\hat \rho \over \tau}
\label{1}
\ee
where the relaxation was considered with respect to the local density operator $\hat \rho^{\rm l.e.}$ or the corresponding local equilibrium distribution function
\be
F(p)=\left [\left ({\varepsilon_0({ p}-{ Q}({R},t))-\mu({R},t)\over
        T({R},t)}\right )+1\right ]^{-1}.
\label{2}
\ee
This local equilibrium is given by a local time dependent 
chemical potential $\mu$,
a local temperature $T$ and a local mass motion momentum $Q$. These
local quantities have been specified by the requirement that the
expectation values for density, momentum and energy are the same as
the expectation values performed with $F$ which ensures conservation laws. 

The correlations are shared in the kinetic equation (\ref{1}) in such
a way that the energy operator $\hat {\cal E}$
parameterizes the density dependent quasiparticle energy or variation of
the energy
functional in the Landau liquid (momentum dependent) or density
functional sense (momentum independent) and the collision integral is
approximated by a conserving relaxation time approximation. While in
\cite{MF99} the general density, energy and momentum dependent form of
such parameterization has been
discussed we want to consider now only a special case of an effective
mass and rigid shift parameterization 
\be
\hat {\cal E}=-\nabla({1\over 2  m(\hat n)})\nabla+ \Delta (\hat n)
\ee
such that the variation ${\cal E}(n)={\cal E}_0+\delta {\cal E}$ 
reads [$p=(p_1+p_2)/2$, $q=p_1-p_2$]
\be
\delta {\cal E} =<p_1| \delta \hat {\cal E}|p_2>=(V_0+V_4 p^2)\delta n(q)
\label{de}
\ee
with
\be
V_0&=&{\delta \over \delta n} \Delta-V_4 {q^2\over 4}+V_q \nonumber\\
V_4&=&{\delta \over \delta n} {1\over 2 m}=-{1\over 2 n m} {\partial
  \ln m\over \partial \ln n}.
\label{V04a}
\ee

The response matrix (\ref{def}) can be given in terms of the
polarization matrix ${\cal P}$, see Eq. (29) of \cite{MF99}, which is the response of the kinetic 
equation without the
self-consistent quasiparticle 
energies $\delta {\cal E}$. The response reads 
\be
{\cal X}=({\cal I}-{\cal P} {\cal G}^{-1} {\cal V})^{-1} {\cal P}
\label{matrix}
\ee
with the matrices simplified for our considered case (\ref{de})
\be
{\cal V}&=&\left (\matrix{g_1 V_0 +g_{p^2} V_4&0&0\cr
g_{\bf p q} V_0+g_{p^2 {\bf pq}} V_4&0&0\cr
g_\epsilon V_0 ++g_{p^2 \epsilon} V_4 &0& 0}\right )_{\omega+{i \over
\tau}}
\label{V}
\\
{\cal G}&=&\left (
\matrix {
g_1         &g_{\bf pq}            & g_\epsilon         \cr
g_{\bf p q} &g_{({\bf p q})^2}     & g_{{\bf p q}\epsilon} \cr
g_\epsilon  &g_{\epsilon {\bf pq}} & g_{\epsilon \epsilon} 
}
\right )_{\omega+{i \over
\tau}}
\label{G}
\ee
and the correlation functions [$\varepsilon=p^2/2m+\Delta$] are
defined as
\be
g_{\phi}(\omega)&=&\int {d p\over (2 \pi)^3} \phi {F({p}+{{q}\over 2})-F({p}-{{q}\over 2}) \over \varepsilon({p}+{{q}\over 2})-\varepsilon({p}-{{q}\over 2})-\omega-i0}.
\label{g}
\ee
Explicit formulae are in appendix~\ref{a}. We keep the matrix notation
of \cite{MF99} also for this special case in order to convince the
reader about the technical usefulness of such notation.
The $3\times 3$ polarization matrix ${\cal P}=\{\Pi_{nm}\}$ contains the 
corresponding density,
momentum and energy polarizations as
\be
&&\left (\matrix {\delta n\cr\delta { q \cdot J}\cr \delta E} \right
)=\left (\matrix{\Pi_{11} \cr \Pi_{21} \cr \Pi_{31}}\right ) \,\,V^{\rm
  ind}.\ee

In fact as found in \cite{MF00} it is possible to express the density
polarization function $\Pi_{11}\equiv \Pi^{\rm n,j,E}$ including density,
current and energy conservation by a simpler one containing density
and energy conservation $\Pi^{\rm n,E}$ as
\be
{1\over \Pi^{\rm n,j,E}(\omega)}-{1\over \Pi^{\rm
      n,E}(\omega)}
&=&{1\over \Pi^{\rm n,j}(\omega)}-{1\over \Pi^{\rm n}(\omega)}=
- {i \omega \over \tau} {m\over n q^2}.
\nonumber\\&&
\label{result}
\ee
This shows that  the momentum conservation leads simply to a dynamical
local field correction. The energy and density conserving polarization
function reads explicitly
\be
\Pi^{\rm n,E}(\omega)&=&(1-i \omega \tau)\left ({g_1(\omega+{i\over \tau}) g_1(0)\over h_1}
\right .\nonumber\\
&&\left . -\omega\tau i {(h_\epsilon g_1(0)-h_1 g_\epsilon(0))^2\over h_1(h_\epsilon^2-h_{\epsilon \epsilon} h_1)}\right )
\label{M-ne}
\ee
where we use the abbreviation
$h_\phi=g_\phi(\omega+{i\over \tau})-\omega \, \tau \, i \,
g_\phi(0)$. The first part is just the known Mermin -Das polarization
function including only density conservation \cite{Mer70,D75}
\be
\Pi^{\rm n}(\omega)&=&\frac{\displaystyle\Pi_0(\omega+{i/\tau})}
 {\displaystyle1-\frac{1}{1-i\omega\tau}\left[1-\frac{\Pi_0(\omega+
             {i/\tau})}{\Pi_0(0)}\right]}\nonumber\\
&=&
(1-i \omega \tau){g_1(\omega+{i\over \tau}) g_1(0)\over h_1}           \label{M-n}.
\ee

Now we want to give the full density response function $\chi={\cal
  X}_{11}$ according to (\ref{def}).
Due to the special considered case (\ref{de}) and consequently (\ref{V}), the
density response function $\chi$ can be written from (\ref{matrix})
into [$\Pi_{11}=\Pi^{\rm n,j,E}$]
\be
\chi(\omega)={\Pi^{\rm n,j,E}(\omega)\over 1-V_0 \Pi^{\rm n,j,E}(\omega) -2 m V_4 \Pi_{13}(\omega)}
\label{chia}
\ee
where
\be
\Pi_{13}(\omega)&=&{\Pi^{\rm n,j,E}\over 2 m}\,\,\,\,
{
\Pi_h \Pi_2(0)-i \tau \omega {\Pi_2(\omega+{i\over \tau})}
\over 
\Pi_h \Pi_0(0)-i \tau \omega {\Pi_0(\omega+{i\over \tau})}
},
\nonumber\\
\Pi_h&=&
{\Pi_2^2(\omega+{i\over \tau})-\Pi_0(\omega+{i\over \tau}) \Pi_4(\omega+{i\over \tau}) \over\Pi_2^2(0)-\Pi_0(0) \Pi_4(0)}
\label{chi1a}
\ee
are expressed in terms of moments of the correlation function (\ref{pn}).
The response function (\ref{chi}) is the main result of this paper
since it gives the consistent response function for the quasiparticle
consisting of effective mass, energy and relaxation time.

\section{Explicit formulae of correlation functions}\label{a}

The different occurring correlation functions (\ref{g}) can be written in terms of moments of the usual Lindhard polarization function $\Pi_0$
\be
\Pi_n=s\int {d {\bf p}\over (2 \pi)^D} p^n{F({\bf p}+{{\bf q}\over 2})-F({\bf p}-{{\bf q}\over 2}) \over {{\bf p q}\over m}-\omega -i0}
\label{pn}
\ee
as
\be
g_1&=&\Pi_0
\nonumber\\
g_{\bf pq}&=&m \omega \Pi_0
\nonumber\\
g_\epsilon&=&{\Pi_2\over 2 m}
\nonumber\\
g_{\epsilon p^2}&=&{\Pi_4 \over 2m}
\nonumber\\
g_{p^2 {\bf pq}}&=&m \omega \Pi_2
\nonumber\\
g_{({\bf pq})^2}&=&- m q^2 {n} + m^2 \omega^2 \Pi_0.
\label{gg}
\ee
Here $s$ is the spin degeneracy and $D$ gives the dimension of the
system. While all formulae in the text are written for the three dimensional
case they hold equally for one - and two dimensions.

For practical and numerical calculations we can rewrite the $\Pi_n$ by 
polynomial division into
\be
\Pi_2&=&-m {n} +{m^2 \omega^2 \over q^2} \Pi_0 +\tilde \Pi_2
\nonumber\\
\Pi_4&=& -{n m q^2 \over 4} (1+{4 m^2 \omega^2 \over q^4})-{m^4 \omega^4 \over q^4} \Pi_0
\nonumber\\
&&-{2 m^2 \omega^2 \over q^2} \tilde \Pi_2-\tilde\Pi_4-{14\over 3} m^2
{E} \left \{\matrix{1\quad {\rm for}\, {D=2,3}\cr
{1\over 2} \quad {\rm for}\, {D=1}}\right .
\label{ei}
\ee
where the $\tilde \Pi_i$ are the projected moments perpendicular to ${\bf q}$
and read
\be
\tilde \Pi_2&=&s \int {d {\bf p}\over (2 \pi)^D} ({\bf p}-{{\bf pq}\over q^2}{\bf q} )^2 {F({\bf p}+{{\bf q}\over 2})-F({\bf p}-{{\bf q}\over 2}) \over {{\bf p q}\over m}-\omega-i0 }\nonumber\\
&=& m\int \limits_{-\infty}^\mu d\mu' \Pi_0
\times \left \{\matrix{2 \quad {\rm for}\, {D=3}\cr
{1} \quad {\rm for}\, {D=1,2}}\right \}
\nonumber\\ 
&\approx& m T \Pi_0 \times \left \{\matrix{2\quad {\rm for}\, {D=3}\cr
{1} \quad {\rm for}\, {D=1,2}}\right \}
\nonumber\\
\tilde \Pi_4&=&s \int {d {\bf p}\over (2 \pi)^D} ({\bf p}-{{\bf pq}\over q^2}{\bf q} )^4 {F({\bf p}+{{\bf q}\over 2})-F({\bf p}-{{\bf q}\over 2}) \over {{\bf p q}\over m}-\omega-i0 }\nonumber\\
&=&m^2\int \limits_{-\infty}^\mu d\mu'\int \limits_{-\infty}^{\mu'}
d\mu'' \Pi_0 
\times \left \{\matrix{8\quad {\rm for}\, {D=3}\cr
{3} \quad {\rm for}\, {D=1,2}}\right \} 
\nonumber\\
&\approx& m^2T^2 \Pi_0 \times \left \{\matrix{8\quad {\rm for}\, {D=3}\cr
{3} \quad {\rm for}\, {D=1,2}}\right \}.
\label{ein}
\ee
The corresponding last identities are valid only for nondegenerate,
Maxwellian, distributions with temperature $T$. The general form of
the polarization functions is presented as an integral over the chemical potential $\mu$ of the Lindhard polarization $\Pi_0$. This is applicable also to the degenerate case.

\subsection{Long wavelength expansion}\label{expans}

In real situations it is often helpful to have the small
wavevector expansion of the various occurring correlation
functions. With the help of (\ref{ein}) and (\ref{ei}) this can be
tremendously simplified if the expansion for $\Pi_0$ is written
\be
\Pi_0(\omega)&=&-{s\over \omega} \int {d p\over (2 \pi)^D} \left (
-{(q \cdot p)^2 \over m^2 \omega} F'+{q^2 (q \cdot p)^2 \over 8 m^3 \omega}
F''\right . \nonumber\\
&&\left .-{(q \cdot p)^4 \over 24 m^4 \omega} F'''-{(p \cdot q)^4 \over m^4 \omega^3}
F'\right )+o(q^5) 
\label{exp3d}
\ee
where $F'=\partial_\mu F$ etc.
For the static case we have
\be
\Pi_0(0)&=&s\int {d p\over (2 \pi)^D} \left (
-F'+{q^2 \over 8 m }
F''
-{(q \cdot p)^2 \over 24 m^2} F'''\right )+o(q^4). 
\nonumber\\&&
\label{exp03d}
\ee
We
give now the one, two and three dimensional case separately.

\subsubsection{3D case}

Since $\partial_p F=-p \partial_\mu F/m$ partial integration gives 
\be
\int {d p\over (2 \pi)^3} F' G(p)=m \int {d p\over (2 \pi)^3} {F\over
  p^2} \partial_p(p G(p))
\label{rule3d}
\ee
and applied to (\ref{exp3d}) one gets
\be
\Pi_0(\omega)={q^2 \over m \omega^2} n+{2 q^4 \over m^2 \omega^4}
E+o({q^6}).
\label{poq}
\ee
The density, $n$, and energy, $E$, and higher moments read in terms of 
\be
f_n={1\over \Gamma(n)} \int\limits_0^\infty {x^{n-1} dx\over {\rm
      e}^{x-\beta \mu}+1}
\ee
as
\be
n=\langle 1 \rangle&=&{s\over \lambda^3} f_{3/2}\nonumber\\
E=\langle {p^2 \over 2 m} \rangle&=&{3\over 2}
\int\limits^\mu d\mu' n={3 \over 2 \beta}{s\over \lambda^3} f_{5/2}\nonumber\\
E_2=\langle \left ({p^2 \over 2 m}\right )^2 \rangle&=&{5\over 2}
\int\limits^\mu d\mu' E={15 \over 4
  \beta^2}{s\over \lambda^3} f_{7/2}
\nonumber\\
E_3=\langle \left ({p^2 \over 2 m}\right )^3 \rangle&=&{7\over 2}
\int\limits^\mu d\mu' E_2={105 \over 8
  \beta^3}{s\over \lambda^3} f_{9/2}.
\nonumber\\&&
\ee
With the help of (\ref{ein}) and (\ref{ei}) one writes down
immediately the higher order correlation functions as 
\be
\tilde \Pi_2(\omega)&=&{4 q^2 \over 3 \omega^2} E+{8 q^4\over 5 m
  \omega^4} E_2 
\nonumber\\
\tilde \Pi_4(\omega)&=&{32 m q^2 \over 15 \omega^2} E_2+{64 q^4\over 35
  \omega^4} E_3.
\ee 
The static case (\ref{exp03d}) yields with (\ref{rule3d})
\be
\Pi_0(0)&=&-\partial_\mu n+{q^2 \over 12 m} \partial_\mu^2 n
\ee
and with (\ref{ei}) and (\ref{ein})
\be
\tilde \Pi_2(0)&=&-2 m n+{q^2 \over 6} \partial_\mu n
\nonumber\\
\tilde \Pi_4(0)&=&-{16\over 3} m^2 E+{2 m q^2 \over 3} n.
\ee 
During the text we use also the small and large wavevector limit of the
static 3-D polarization functions.
Since
\be
\Pi_0(0)=\left \{ \matrix{-n^2 {\cal K}_0+o(q^2)
\cr -4 n{m\over q^2}-{32 m^2\over 3 q^4} E+o(1/q^6)} \right .
\label{ex0}
\ee
we obtain from (\ref{ei}) and (\ref{ein})
\be
\Pi_2(0)=\left \{ \matrix{-3 mn +o(q^2)\cr
-m n -{16\over 3} {m^2 \over q^2} E-{128 m^3\over 15 q^4} E_2+o(1/q^6)}\right .
\label{ex2}
\ee
which leads to 
\be
{\Pi_2(0)\over \Pi_0(0)}=\left \{\matrix{{3 m\over n {\cal K}_0}+o(q^2)\cr
    {q^2\over 4}+{2 E m\over 3 n}+o(1/q^2)}\right ..
\label{b16}
\ee

\subsubsection{2D case}

For two dimensions we have instead of (\ref{rule3d})
\be
\int {d p\over (2 \pi)^2} F' G(p)=m \int {d p\over (2 \pi)^2}
F\partial_p G(p)
\label{rule2d}
\ee
which applied to (\ref{exp3d}) yields
\be
\Pi_0(\omega)={2 \pi q^2 \over  m \omega^2} n+{6 \pi q^4 \over  m^2 \omega^4} E
\ee
and from (\ref{ei}) and (\ref{ein})
\be
\tilde \Pi_2(\omega)&=&{2 \pi q^2 \over  \omega^2} E+{6 \pi q^4 \over  m \omega^4} E_2
\nonumber\\
\tilde \Pi_4(\omega)&=&{6 \pi q^2 m\over \omega^2} E_2+{18 \pi q^4 \over   \omega^4} E_3.
\ee
The different occurring moments reads here
\be
n=\langle 1 \rangle&=&{s\over \pi \lambda^2} f_{1}\nonumber\\
E=\langle {p^2 \over 2 m} \rangle&=&
\int\limits^\mu d\mu' n={1 \over \beta}{s\over \lambda^2} f_{2}\nonumber\\
E_2=\langle \left ({p^2 \over 2 m}\right )^2 \rangle&=&
\int\limits^\mu d\mu' E={1 \over 
  \beta^2}{s\over \lambda^2} f_{3}
\nonumber\\
E_3=\langle \left ({p^2 \over 2 m}\right )^3 \rangle&=&
\int\limits^\mu d\mu' E_2={1\over 
  \beta^3}{s\over \lambda^2} f_{4}.
\nonumber\\&&
\ee
The static case is now analogously and reads with (\ref{rule2d}) from
(\ref{exp03d})
\be
\Pi_0(0)&=&-\partial_\mu n+{ q^2 \over 24 m} \partial_\mu^2 n
\nonumber\\
\tilde \Pi_2(0)&=&-m n+{ q^2 \over 24 } \partial_\mu n
\nonumber\\
\tilde \Pi_4(0)&=&-3 m^2 E+{ m q^2 \over 8 } n.
\ee

\subsubsection{1D case}

For one dimensions we have instead of (\ref{rule3d})
\be
\int {d p\over (2 \pi)} F' G(p)=m \int {d p\over (2 \pi)} F
\partial_p({G(p)\over p})
\label{rule1d}
\ee
which applied to (\ref{exp3d}) yields
\be
\Pi_0(\omega)&=&{q^2 \over 3 m \omega^2} n-{q^4\over 60 m^2 \omega^2}
\partial_\mu n+{6 q^4 \over 5 m^2 \omega^4} E
\nonumber\\
\tilde \Pi_2(\omega)&=&{2 q^2 \over 3 \omega^2} E-{q^4\over 60 m \omega^2}
n+{4 q^4 \over 5 m \omega^4} E_2
\nonumber\\
\tilde \Pi_4(\omega)&=&{4 m q^2 \over 3 \omega^2} E_2-{q^4\over 10  \omega^2}
E+{24 q^4 \over 25  \omega^4} E_3.
\ee
The static case is 
\be
\Pi_0(0)&=&-\partial_\mu n+{q^2 \over 12 m} \partial_\mu^2 n
\nonumber\\
\tilde \Pi_2(0)&=&-m n+{ q^2 \over 12 } \partial_\mu n
\nonumber\\
\tilde \Pi_4(0)&=&-=-6 m^2 E+{ m q^2 \over 4 } n.
\ee
The occurring moments reads here
\be
n=\langle 1 \rangle&=&{s\over 4 \lambda} f_{1/2}\nonumber\\
E=\langle {p^2 \over 2 m} \rangle&=&{1\over 2}
\int\limits^\mu d\mu' n={1 \over 8 \beta}{s\over \lambda} f_{3/2}\nonumber\\
E_2=\langle \left ({p^2 \over 2 m}\right )^2 \rangle&=&{3\over 2}
\int\limits^\mu d\mu' E={3 \over 16
  \beta^2}{s\over \lambda} f_{5/2}
\nonumber\\
E_3=\langle \left ({p^2 \over 2 m}\right )^3 \rangle&=&{5\over 2}
\int\limits^\mu d\mu' E_2={15 \over 32
  \beta^3}{s\over \lambda} f_{7/2}.
\nonumber\\&&
\ee

\subsection{Large frequency limit}

The large frequency limit can be given analogously to the forgoing
chapter. We restrict here to give the expansion for 3D
\be
\Pi_0(\omega)={q^2 \over m \omega^2} n+{2 q^4 \over m^2 \omega^4}
E+{q^6\over 4 m^3 \omega^4} n +o({1\over \omega^6}).
\label{pow}
\ee
Please remark the difference to the large wavelength expansion
(\ref{poq}). The corresponding higher order correlation functions are
completely analogously given by the methods of the foregoing chapter.
With (\ref{ein}) and (\ref{ei}) one gets
\be
\tilde \Pi_2(\omega)&=&{4 q^2 \over 3 \omega^2} E+{8 q^4\over 5 m
  \omega^4} E_2 +{q^6 \over 3 m^2 \omega^4}E+o({1\over \omega^6})
\nonumber\\
\tilde \Pi_4(\omega)&=&{32 m q^2 \over 15 \omega^2} E_2+{64 q^4\over 35
  \omega^4} E_3+{8 q^6 \over 15 m \omega^4} E_2+o({1\over \omega^6}).
\nonumber\\&&
\label{p24w}
\ee

\section{Perturbation theory and frequency sum rules for 1,2,3 dimensions}\label{pert}

The external potential is adiabatically switched on 
\be
V^{\rm
  ext}(r,t)=V(r) {\rm e}^{0 t} \Theta (-t)
\ee
and induces a time dependent change in the Hamilton operator
\be
\delta \hat H(t)=-\int dr \hat n(r,t) V^{\rm ext} (r,t).
\ee
The variation of the density matrix operator $\hat \rho(t)=\hat \rho+\delta
\hat \rho(t)$ can be
found from the linearized van--Neumann equation as
\be
\delta \hat \rho(t)=-i\int \limits_{-\infty}^t [\delta \hat H,\hat \rho_0]
\ee
where it has been assumed that the perturbation is conserving
symmetries of the equilibrium Hamiltonian $[\hat H_0,\delta \hat \rho]=0$.

The variation of the density expectation value $\delta n={\rm Tr} \delta
\rho\, \hat n$ is consequently 
\be
&&\delta n(r,t)=i \int\limits_{-\infty}^t dt' \int dr' V(r',t') \langle[\hat
n(r,t),\hat n(r',t')]\rangle.
\nonumber\\&&
\label{n1}
\ee
Since in equilibrium the commutator is only dependent on the
difference of coordinates and times we can define
\be
-2 {\rm Im} \chi(q,\omega)&=&\int dt {\rm e}^{i \omega (t-t')}\int d r
d r' {\rm e}^{-i q(r-r')}\nonumber\\&&\times\langle[\hat n(r,t),\hat n(r',t')]\rangle
\label{im}
\ee
from which we obtain the Fourier transform of (\ref{n1}) to
\be
\delta n(q,\omega)=V^{\rm ext}(q,\omega) \int {d {\bar \omega} \over \pi} {{\rm Im } \chi
  (q,{\bar \omega})\over {\bar \omega}-\omega -i 0}
\label{ch}
\ee
where $V^{\rm ext}(q,\omega)=V^{\rm ext}(q)/(0+i \omega)$. This is of
course identical with (\ref{dela}).

\subsection{Sum rules}

Now one can derive the first and second order sum rules of the imaginary
part of the response function (\ref{im}). Therefore we generalize the
definition (\ref{im}) to nonequilibrium and finite
systems
\be
&&{\rm Im} \chi(q,\omega,R,t)=-\frac 1 2 
\int d\tau {\rm e}^{i \omega \tau}\int d r
 {\rm e}^{-i q r}\nonumber\\&&\times\langle[\hat n(R+r/2,t+\tau/2),\hat n(R-r/2,t-\tau/2)]\rangle.
\label{im1}
\ee
Higher order moments can be expressed by correlation functions as well
\cite{MM88}. Here we restrict to the lowest two orders and rederive it
in conventional way.

If we assume further equilibrium but finite systems we can define an
averaged response by applying spatial averaging $\int dR/V$ to
(\ref{im1}) such that we obtain
\be
{\rm Im} \chi(q,\omega)&=&\!-\!\frac {1}{2V} 
\int d\tau {\rm e}^{i \omega t}\langle[\hat n(q,t),\hat n(-q,0)]\rangle.
\label{im2}
\ee
From this expression it is easy to see that the first two frequency sum
rules read
\be
\int {d \omega \over \pi} \omega {\rm Im} \chi (q,\omega)&=&-\frac 1 V
\langle[i\partial_t \hat n(q,t)|_{t=0},\hat n(-q,0)]\rangle
\nonumber\\
\int {d \omega \over \pi} \omega^3 {\rm Im} \chi (q,\omega)\!&=&\!\!-\!\frac 1 V
\langle[(i \partial_t)^3 \hat n(q,t)|_{t\!=\!0},\hat n(-q,0)]\rangle.
\label{n2}
\nonumber\\&&
\ee
Using the Heisenberg equation $i\partial_t \hat n=[\hat n,\hat H]$
and
\be
\hat H&=&\int {dp \over (2 \pi)^D} {p^2\over 2 m} \hat a_p^+\hat a_p
\nonumber\\&&+\frac 1
2 \int {dp dp_1 dp_2\over (2 \pi)^{3D}} V_p
\hat a_{p_1}^+\hat a_{p_2}^+\hat a_{p_2+p}\hat a_{p_1-p}
\nonumber\\
\hat n_q&=&\int {dp \over (2 \pi)^D} \hat a_p^+\hat a_{p+q}
\ee
we
can express the sum rules (\ref{n2}) as
\be
\int {d \omega \over \pi} \omega {\rm Im} \chi (q,\omega)&=&\frac 1 V
\langle[q\hat j_q,\hat n_{-q}]\rangle
\nonumber\\
\int {d \omega \over \pi} \omega^3 {\rm Im} \chi (q,\omega)&=&-\frac 1 V
\langle[q \hat j_q,[q \hat j_{-q},\hat H]]\rangle
\label{n3}
\ee
where the divergence of the current operator reads
\be
q \hat j_q=\int {dp \over (2 \pi)^D} {2 p q +q^2 \over 2 m}\hat
a_p^+\hat a_{p+q}.
\ee   
Performing the last commutators one obtains finally
\be
\int {d \omega \over \pi} \omega {\rm Im} \chi (q,\omega)&=&{q^2 \over
  m} {\langle\hat n_{q=0}\rangle\over V}=n {q^2 \over m}
\ee
and
\be
&&\int\!\! {d \omega \over \pi} \omega^3 {\rm Im} \chi (q,\omega)\!=\!
{n q^6\over 4 m^3}\!+\!{3 q^2 \over m^3 V}\!\!\int \!\! \!{dp \over (2
  \pi)^D} (p \cdot q)^2 \langle\hat a_p^+\hat a_{p}\rangle
\nonumber\\
&& 
{1\over m^2 V}\!\!\int\!\!\! {dp \over (2 \pi)^D} V_p\!
\left \{
(p \cdot q)^2 \!\!\int\!\!\! {dp_1 dp_2\over (2 \pi)^{2D}}
\hat a_{p_1}^+\hat a_{p_2}^+\hat a_{p_2+p+q}\hat a_{p_1-p-q}
\right .
\nonumber\\
&&\left .-((p \cdot q)^2-q^2 (p \cdot q)) \int {dp_1 dp_2\over (2 \pi)^{2D}}
\hat a_{p_1}^+\hat a_{p_2}^+\hat a_{p_2-p}\hat a_{p_1+p}
\right \}.
\nonumber\\&&
\label{n4}
\ee
Since we had symmetric expressions $\phi(p)=\phi(-p)$
the second term leads just to kinetic energy density
\be
&&{3 q^2 \over m^3 V}\int {dp \over (2
  \pi)^D} (p \cdot q)^2 \langle\hat a_p^+\hat a_{p}\rangle
\nonumber\\&&=
{2 q^4 \over m^2 V}\int {dp \over (2
  \pi)^D} {p^2\over 2 m} \langle\hat a_p^+\hat a_{p}\rangle
\times
\left\{\matrix{3\quad {\rm for}\, 1D\cr 1\quad {\rm for}\, 2,3 D}\right . 
\nonumber\\&&
={2 q^2 E \over m^2}
\times
\left\{\matrix{3\quad {\rm for}\, 1D\cr 1\quad {\rm for}\, 2,3 D}\right .. 
\ee

Now we are going to express the last 4 creation and annihilation
operators by the structure function itself. Therefore we use the
definition of the pair correlation function
\be
\langle\hat a_{r_1}^+\hat a_{r_2}^+\hat a_{r_2}\hat a_{r_1}\rangle=g_r n(r_1)
n(r_2).
\ee
Applying the spatial averaging $\int d R/V$ where $R=(r_1+r_2)/2$ and
Fourier-transform the difference $r_1-r_2$ into $q$ we obtain
\be
&&\int {dp_1 dp_2\over (2 \pi)^{2D} V}
\hat a_{p_1}^+\hat a_{p_2}^+\hat a_{p_2+q}\hat a_{p_1-q}
\nonumber\\&&
=
\int dr {\rm e}^{-i r \cdot q} g_r \int {d R\over V} n(R+r/2) n(R-r/2)
\nonumber\\&&
=n^2  \int dr {\rm e}^{-i r \cdot q} (g_r-1)+n^2 (2 \pi)^3 \delta(q)
\nonumber\\&&
=n (S_q-1)+n^2 (2 \pi)^3 \delta(q)
\label{b17}
\ee
where we neglected spatial gradients in the density and used (\ref{gs})
for the last step.
Using (\ref{b17}) in (\ref{n4}) we obtain finally
\be
\int {d \omega \over \pi} \omega^3 {\rm Im} \chi (q,\omega)&=&
{n q^6\over 4 m^3}+{2 q^2 E \over m^2}\times
\left\{\matrix{3\quad {\rm for}\, 1D\cr 1\quad {\rm for}\, 2,3 D}\right \} 
\nonumber\\&-&{n^2 \over m^2}q^4 V_q \left (\tilde I(q)+\left\{\matrix{\tilde I_1(q)\quad {\rm for}\, 1D\cr 0\quad {\rm for}\, 2,3 D}\right \} 
\right )
\nonumber\\&&
\ee
with (\ref{I})
\be
&&\tilde I(q)\!=\!-{1\over n} \int {d k \over (2 \pi)^D} (S_{k-q}\!-\!S_k \!+\!n \delta_{k,q}\!-\!n\delta_{k,0})
  {(k\cdot q)^2\over q^4} {V_k\over V_q}
\nonumber\\&&
\label{I1}
\ee
where we understand $\delta_k=(2\pi)^3 \delta(k)$.
For the one-dimensional case an extra term appears
\be
\tilde I_1(q)={1\over n} \int {d k \over (2 \pi)} (S_k -1+n \delta_{k,0})
  {k V_k\over q V_q}.
\label{I11}
\ee

In the following we restrict to the required formulae for the three
dimensional case. One can  Fourier transform
\be
\tilde I(q)=-\int dr g_r(1-\cos{(q \cdot r)}) {(q \cdot \partial_r)^2 V_r\over q^4 V_q}
\label{Puff}
\ee
which was first given by Puff \cite{P65}. This correct form leads
unavoidably to the appearance of the $\delta_p$ terms in (\ref{I1})
very often overseen in later papers. For asymptotic expansions,
however, we have to be careful that $g_r-1$ is the object which
renders spatial integrals finite. Therefore the $\delta_p$ terms in 
(\ref{I1}) have to be considered separately
\be
-{1\over n} \int {d k \over (2 \pi)^3} (\!n \delta_{k,q}\!-\!n\delta_{k,0})
  {(k\cdot q)^2\over q^4} {V_k\over V_q}=-1
\ee
such that we obtain instead of (\ref{Puff})
\be
&&\tilde I(q)=-\int dr (g_r-1)(1-\cos{(q \cdot r)}) {(q \cdot
  \partial_r)^2 V_r\over q^4 V_q}-1.
\nonumber\\&&
\label{Puff1}
\ee
For Coulomb potentials we can further simplify
\be
\tilde I(q)&=&-2 \int \limits_0^{\infty} {d r\over r} (g_r-1) j_2(q r)-1
\label{cpot}
\ee
with the spherical Bessel function $j_2(x)$.
From this expression one sees the small wavevector limit
\be
\tilde I(q)&=&-{2 q^2 \over 15}  \int \limits_0^{\infty} {d r} r
(g_r-1)-1 +o(q^4)
\nonumber\\
&=&\frac 2 5 {q^2 \over k_f^2} \gamma -1 +o(q^4)
\label{small}
\ee
where we have used (\ref{gs}) and the definition (\ref{gamma}).
The long wavevector limit takes the form
\be
\tilde I(q)&=&-2 \int \limits_0^{\infty} {d x\over x} (g_{x/q}-1) j_2(x)-1
\nonumber\\
&=&-2 (g_0-1) \int \limits_0^{\infty} {d x\over x} j_2(x)-1+o(1/q^2)
\nonumber\\&=&\frac 2 3 (1-g_0)-1+o(1/q^2).
\label{l}
\ee

\end{document}